\documentclass[prl,footinbib,twocolumn,showpacs,amsmath,amssymb,reprint]{revtex4-1}
\usepackage{graphicx,amsmath,bm}
\usepackage{color}

\begin{document}
\draft
\title{Nonequilibrium critical Casimir interactions in binary fluids}

\author{Akira Furukawa$^1$, Andrea Gambassi$^2$, Siegfried Dietrich$^3$, and Hajime Tanaka$^1$ } 

\affiliation{$^1$Institute of Industrial Science, University of Tokyo, Meguro-ku, Tokyo 153-8505, Japan\\
$^2$SISSA  -- International School for Advanced Studies and INFN, via Bonomea 265,  34136 Trieste,
Italy\\
$^3$Max-Planck-Institut f\"ur Intelligente Systeme, Heisenbergstr.~3, 70569 Stuttgart, Germany and \\ Institut f\"ur Theoretische Physik IV, Universit\"at Stuttgart, Pfaffenwaldring 57, 70569 Stuttgart, Germany}

\date{Received February 26, 2013}

\begin{abstract}
Colloids immersed in a critical binary liquid mixture are subject to critical Casimir forces (CCFs) because they confine its concentration fluctuations and influence the latter via effective surface fields. To date, CCFs have only been studied in thermodynamic equilibrium. However, due to the critical slowing down, the order parameter around a particle can easily be perturbed by any motion of the colloid or by solvent flow. This leads to significant but largely unexplored changes in the CCF. Here we study the drag force on a single colloidal particle moving in a near-critical fluid mixture and the relative motion of two colloids due to the CCF acting on them. In order to account for the kinetic couplings among the order parameter field, the solvent velocity field, and the particle motion, we use a fluid particle dynamics method. These studies extend the understanding of CCFs from thermal equilibrium to non-equilibrium processes, which are relevant to current experiments, and 
show the emergence of significant effects near the critical point.  
\end{abstract}

\maketitle

Upon approaching a bulk critical point $T_{\rm c}$ of demixing in a critical binary liquid mixture, both the bulk correlation length $\xi$ and the lifetime $\tau_\xi$ of the order parameter fluctuations --- given by the deviation $\psi(\bm{r})$ of the local concentration from the bulk critical concentration --- diverge \cite{hohenberg1977theory,onuki2002}: 
$\xi =\xi_{0,\pm} \epsilon^{-\nu}$ and $\tau_\xi=\tau_{0,\pm} \epsilon^{-\nu z}$ for 
$\epsilon=(T-T_{\rm c})/T_{\rm c} \rightarrow 0^{\pm}$ with the  
bulk static and dynamic critical exponents $\nu$ and $z$, respectively.
When colloidal particles are suspended in such a critical binary liquid mixture, preferential adsorption 
of one of its components at the colloid surfaces 
effectively is tantamount to  
an ordering boundary condition for $\psi$. 
The fluctuation-induced and surface-field supported slow spatial decay of $\psi$ 
occurs on the scale of $\xi$ and leads to long-ranged effective interactions between the 
colloids \cite{fisher1978,kardar1999,dietrich1998,dietrich2003,bechinger2008,okamoto_onuki,gambassi2009casimir}. 
The same phenomenon has also been observed in wetting films of a classical binary liquid mixture \cite{fukuto2005critical}, 
of $^4$He \cite{garcia1999critical,ganshin2006critical},
and of $^4$He/$^3$He \cite{garcia2002} near the superfluid transition. 
These effective interactions are referred to as CCFs, 
in view of the analogy with the corresponding quantum force \cite{casimir,casimirR2009,kardar1999,lamoreaux2005}, 
which is caused by the confinement of the vacuum 
fluctuations of the electromagnetic field. 
There is satisfactory agreement between the experimental 
data \cite{garcia1999critical,garcia2002,ganshin2006critical,fukuto2005critical,bechinger2008} 
and the wealth of the corresponding theoretical predictions (see, e.g., Refs.~\cite{gambassi2009casimir,gambassi2009critical} and references therein). 

So far CCFs have been studied primarily in thermodynamic equilibrium 
on the basis of finite-size scaling of the free energy \cite{indekeu1985,krech1991}. 
However, since both $\xi$ and $\tau_\xi$  diverge at 
the critical point, any motion of the colloidal particles --- even the one induced by CCFs themselves ---  
perturbs significantly the static situation. 
Understanding this issue requires one to analyze critical dynamics in confinement by accounting for the
dynamical couplings between $\psi$, the velocity field $\bm{v}$ 
of the liquid mixture, and the particle motion, which were considered only partially in previous 
studies~\cite{gambassi2006critical,gambassi2008relaxation,Dean2010,Dean2010b,Dean2010c}.  
Since at present an analytic solution for this problem is beyond reach, we simulate these couplings by using a fluid particle dynamics (FPD) method \cite{FPD}. 
Interesting time-dependent properties associated with the CCF 
include the two-point correlation function of the instantaneous force at different times, the relaxation of the CCF  
upon a thermodynamic quench, shear effects on the CCF, 
the drag force on a particle pulled through the solvent, or the time dependence of the distance between two colloids approaching their equilibrium configuration. Here we focus on the last two issues.

\emph{Method.---}Within FPD \cite{FPD} a set of $N$ particles with centers of mass located at $\bm{r}_i$, $i=1,\ldots,N$, is
represented by a spatially varying 
viscosity $\eta(\bm{r})=\eta_{\ell}+\sum_{i=1}^{N}(\eta_{\rm c}-\eta_{\ell})\phi_i(\bm{r})$, where $\eta_{\ell}$ and $\eta_{\rm c}$ are the viscosities of the liquid solvent and of a ``fluid'' mimicking a colloid, respectively. $\phi_i$ describes the $i$-th particle as $\phi_i({\bm{r}})=\{1+\tanh [(a-|{\bm{r}}-{\bm{r}}_i|)/d]\}/2$, where $a$ is the particle radius and $d$ its interface thickness.
The key points of FPD are that 
(i) the particle rigidity is maintained approximately by the large viscosity 
difference, $(\eta_{\rm c}-\eta_\ell)/\eta_\ell\gg 1$, and 
(ii) the in fact sharp solid-liquid boundary is replaced by a smooth interface 
with thickness $d$. 
Upon decreasing $d$ and increasing $\eta_{\rm c}/\eta_{\ell}$ 
the method is expected to model actual colloids \cite{FPD,tanaka_review_2006,furukawa2010,fujitani2006}.  
The free energy functional for the colloidal suspension is given by 
$F\{\psi,\bm{r}_i\} = F_{\rm b}\{\psi\}+F_{\rm s}\{\psi,\bm{r}_i\}+F_{\rm a}\{\psi,{\bm{r}}_i\} +U\{\bm{r}_i\}$. 
The first term is the bulk free energy $F_{\rm b}\{\psi\} =\int {\rm d}^3r\, [f_{\rm b}(\psi)+{K_0}(\nabla\psi)^2/2]$ with the Ginzburg-Landau form 
$f_{\rm b}(\psi) = r_0\epsilon\psi^2/2+u_0 \psi^4/4$
at the bulk critical concentration; 
$r_0$, $u_0$, and $K_0$ are positive constants such that
$\xi_{0,+}=\sqrt{K_0/r_0}$, $\nu=\frac{1}{2}$, and $\psi_b(\epsilon \to 0^-)=\psi_0(-\epsilon)^\beta$
with $\beta=\frac{1}{2}$ and $\psi_0=\sqrt{r_0/u_0}$. 
The second term represents the quasi-surface contribution   
$F_{\rm s}\{\psi,\bm{r}_i\} =\sum_{i=1}^{N}\int {\rm d}^3r\, |\nabla \phi_i|h_1 \psi$,  
with the value of the surface field $h_1$ depending on the details of the actual interaction 
between the colloid surface and the two species of the solvent. 
The third term $F_{\rm a}\{\psi,{\bm{r}}_i\} =\sum_{i=1}^{N}\int {\rm d}^3r\,
\chi \phi_i \psi^2$
with $\chi>0$ is a coupling introduced to suppress $\psi$ 
inside each particle.  
Finally, the potential energy $U\{\bm{r}_i\}=\sum_{i> j}u(|\bm{r}_i-\bm{r}_j|)$ accounts for the direct pair potential 
$u(\bm{r})$ between the colloids.

The dynamic equations can be constructed on the basis of the standard model H \cite{hohenberg1977theory,onuki2002} for critical dynamics of a binary liquid mixture. 
The velocity field $\bm{v}(\bm{r})$ of the fluid, which 
consists of both the solvent and the colloids, 
obeys the Navier-Stokes equation 
\begin{equation}
\rho \left( \partial_t
+\bm{v}\cdot\nabla\right)\bm{v} = {\bm{f}}_{\rm rev}-\nabla\cdot{\stackrel{\leftrightarrow}{\mbox{\boldmath$\sigma$}}}, 
\label{Navier-Stokes}
\end{equation}
with $\stackrel{\leftrightarrow}{\mbox{$\bm\sigma$}} = p\!\!\stackrel{\leftrightarrow}
{\mbox{$\bm I$}}-\eta((\nabla{\bm{v}})^\dagger+\nabla{\bm{v}})$
where $\stackrel{\leftrightarrow}{\mbox{$\bm I$}}$ is the unit tensor. The
hydrostatic pressure $p$ facilitates to satisfy the incompressibility condition
$\nabla\cdot {\bm{v}}=0$ and ${\bm{f}}_{\rm rev}({\bm{r}})$ 
is the reversible force density (see below).
We consider the mass density $\rho$ of the liquid to be spatially constant and thus to be the same as the one of the particles;
$\psi(\bm{r})$ satisfies 
\begin{equation}
\partial_t \psi = -\nabla\cdot {\mbox{\boldmath$j$}}_\psi,  
\label{continuity}
\end{equation}
where 
$\bm{j}_\psi = \psi{\bm{v}}-{L_\psi\nabla({\delta F}/{\delta \psi})}$
is the flux of $\psi$ with the kinetic coefficient $L_\psi$. 
For convenience, we adopt a spatially varying $L_\psi=(1-\sum_{i=1}^N\phi_i)L$,   
with constant $L>0$, which vanishes inside each colloid where 
$\phi_i=1$.
Together with $F_{\rm a}$, this guarantees that inside the particles there is no diffusive flux so that the concentration  
remains unchanged, i.e., $\psi({\bm{r}},t)=0$, as it should be inside solid particles. 
Within the FPD method, the velocity $\bm{V}_i$ of the $i$-th fluid particle is given by 
\begin{equation}
\bm{V}_i =\dfrac{1}{\Omega_i}\int_{{\mathbb R}^3}\!\! {\rm d}^3r\, \phi_i {\bm{v}}, \quad
\quad  \Omega_i= \int_{{\mathbb R}^3} \!{\rm d}^3r\, \phi_i.
\label{particle_velocity}
\end{equation}
In the presence of a smooth interface, 
the standard expression \cite{onuki2002} for the reversible force density ${\bm{f}}_{\rm rev}$ has an additional contribution \cite{furukawa_method};  
from $F\{\psi,\bm{r}_i\}$ and the dynamic equations 
(\ref{Navier-Stokes}), (\ref{continuity}), and (\ref{particle_velocity})  
one obtains 
\begin{equation}
{\bm{f}}_{\rm rev} =
-\psi\nabla\dfrac{\delta F}{\delta \psi} -\sum_{i=1}^{N}\dfrac{\phi_i}{\Omega_i}\dfrac{\partial F}{\partial \bm{r}_i}, 
\label{reversible_force}
\end{equation}
where the first term contains  
the usual bulk osmotic force $-\psi \nabla ({\delta F_{\rm b}}/{\delta \psi})$
due to $F_{\rm b}$, 
which can be expressed as $-\nabla\cdot {\stackrel{\leftrightarrow}{\mbox{$\bm{\Pi}$}}}$  in terms of 
the osmotic pressure tensor ${\stackrel{\leftrightarrow}{\bm{\Pi}}} =  ( \psi{\delta F_{\rm b}}/{\delta \psi} - f_{\rm b} -{K_0}(\nabla\psi)^2/2 ){\stackrel{\leftrightarrow}{\bm{I}}}+K_0 \nabla\psi\nabla\psi$ \cite{onuki2002}.
As it will be detailed elsewhere \cite{furukawa_method}, one can show that 
the contributions of $F_{\rm s}$ and $F_{\rm a}$ to Eq.~(\ref{reversible_force})  vanish 
for $d/a \rightarrow 0$.

In the actual numerical calculations, the velocity field
${\bm{v}}({\bm{r}},t+\Delta t)$ is obtained on the lattice from 
$\bm{v}$, $\psi$, and ${\bm r}_i$
at time $t$ 
via Eq.~(\ref{Navier-Stokes}).  
The particles are then moved off-lattice according to 
${\bm{r}}_i(t+\Delta t)={\bm{r}}_i(t)+\Delta t\, \bm{V}_i(t+\Delta t)$ (see Eq.~(\ref{particle_velocity})), 
where $\Delta t$ is the time increment of the numerical integration, and then put back on the lattice. 
This procedure is repeated for each time step $\Delta t$. 

In our simulation, 
we have used $d$ as the unit of length (and as the lattice constant) 
and $\tau=\rho d^2/\eta_\ell$ as the unit of time, which render
the scaled $\rho$ and $\eta_\ell$ to be 1. 
The units of stress and energy are 
$\bar\sigma=\rho(d/\tau)^2$ and $\bar{E}=\bar{\sigma} d^3$, 
respectively. 
In units of $d^2/(r_0\tau)$, $L$ scales to 1.
In the free energy functional 
we set $r_0/\bar{\sigma}=u_0/r_0=K_0/(d^2 r_0)=1$, $h_1/(d r_0)=-10$ or $-4$, 
and $\chi/r_0=2$, 
which fix the bulk non-universal amplitudes $\xi_{0,+}=d\; (=1)$ and $\psi_0=1$. 
Furthermore, we set $\eta_{\rm c}/\eta_{\ell}=50$ and $\Delta t=0.003$.  
In the following all variables are understood as scaled. 
The translation of the above choices for the dimensionless quantities (motivated by computing limitations)
into physical ones results into very short time and length scales. 
However, the phenomena we focus on here are caused primarily by the mismatch between 
the time scales for concentration diffusion and advection, which we shall quantify and compare with experimentally relevant figures below. 
In this study we focus on the critical composition with average concentration $\bar{\psi}=0$. 
Although the CCF is known to increase upon moving away from this composition \cite{dietrich1998,dietrich2003,okamoto_onuki},
we focus on the critical one because it is relevant for current experiments and 
correspondingly, $\xi$ is maximized, 
resulting in enhanced dynamical effects, which in fact become more severe as $\xi$ grows (see below). 
We solve the equation of motion [Eq.~(\ref{Navier-Stokes})] by the marker-and-cell method \cite{harlow1965numerical}. 
The present mean-field analysis neglects the effects of thermal fluctuations, which are left to future studies.  

\begin{figure}
\begin{center}
\includegraphics[width=0.40\textwidth]{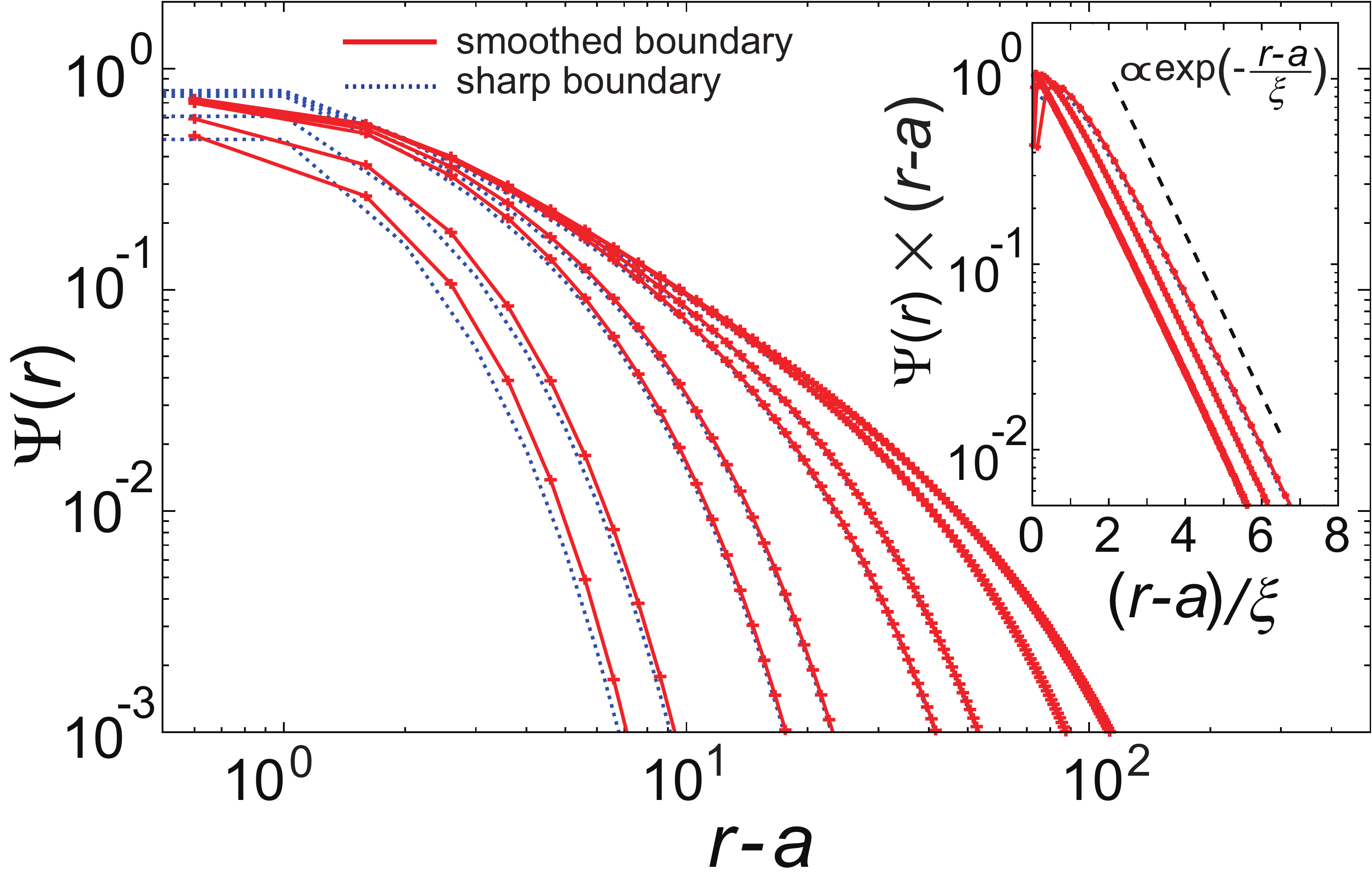}
\end{center}
\caption{(color online) 
Equilibrium concentration profile $\psi$ around a single particle of radius $a=6.4$ for $\epsilon=$1, 0.5, 0.1, 0.05, 0.01, 0.005, 0.001, and 0.0005
from bottom to top, with $h_1=-4$ and as a function of the distance from the surface of the particle. 
Upon approaching the critical point, i.e., for $\epsilon \lesssim 0.1$ the difference between smoothed (solid line) and sharp boundaries (dotted line, see main text) is visible only within a distance $d\; (=1)$ from the surface.
For $\epsilon=0.05$, 0.005, and 0.0005, the log-linear plot in the inset shows agreement with the theoretically predicted exponential decay (dashed line).}
\label{Fig1}
\end{figure} 

\emph{Equilibrium.---}In order to assess the reliability of FPD  
we first consider relaxation towards equilibrium for a single particle.
In the absence of a mean velocity field, the system is spherically symmetric 
around the center $r=0$ of the particle.
For various reduced temperatures, Fig.~\ref{Fig1} compares the  
equilibrium $\psi(r)$, 
as obtained by evolving an arbitrarily chosen initial condition with FPD (smooth boundary, solid line), with the one (dashed line) obtained  by imposing at the sharp boundary $r=a$ of the particle appropriate conditions \cite{fukuda2006} for $\psi$ and $\bm{v}$  during their evolution with model H.
Due to $h_1<0$, the critical adsorption layer is formed with $\psi>0$ and
its thickness is given by  $\xi = \epsilon^{-1/2}$, as highlighted in the inset. 
Figure~\ref{Fig1} demonstrates  
that the two sets of profiles are indistinguishable, apart from distances $r-a \lesssim d =1$ 
which are outside the scaling regime. 
This illustrates that FPD reproduces accurately the mean-field equilibrium concentration profile except in the close vicinity of the surface.

\begin{figure}
\begin{center}
\includegraphics[width=0.47\textwidth]{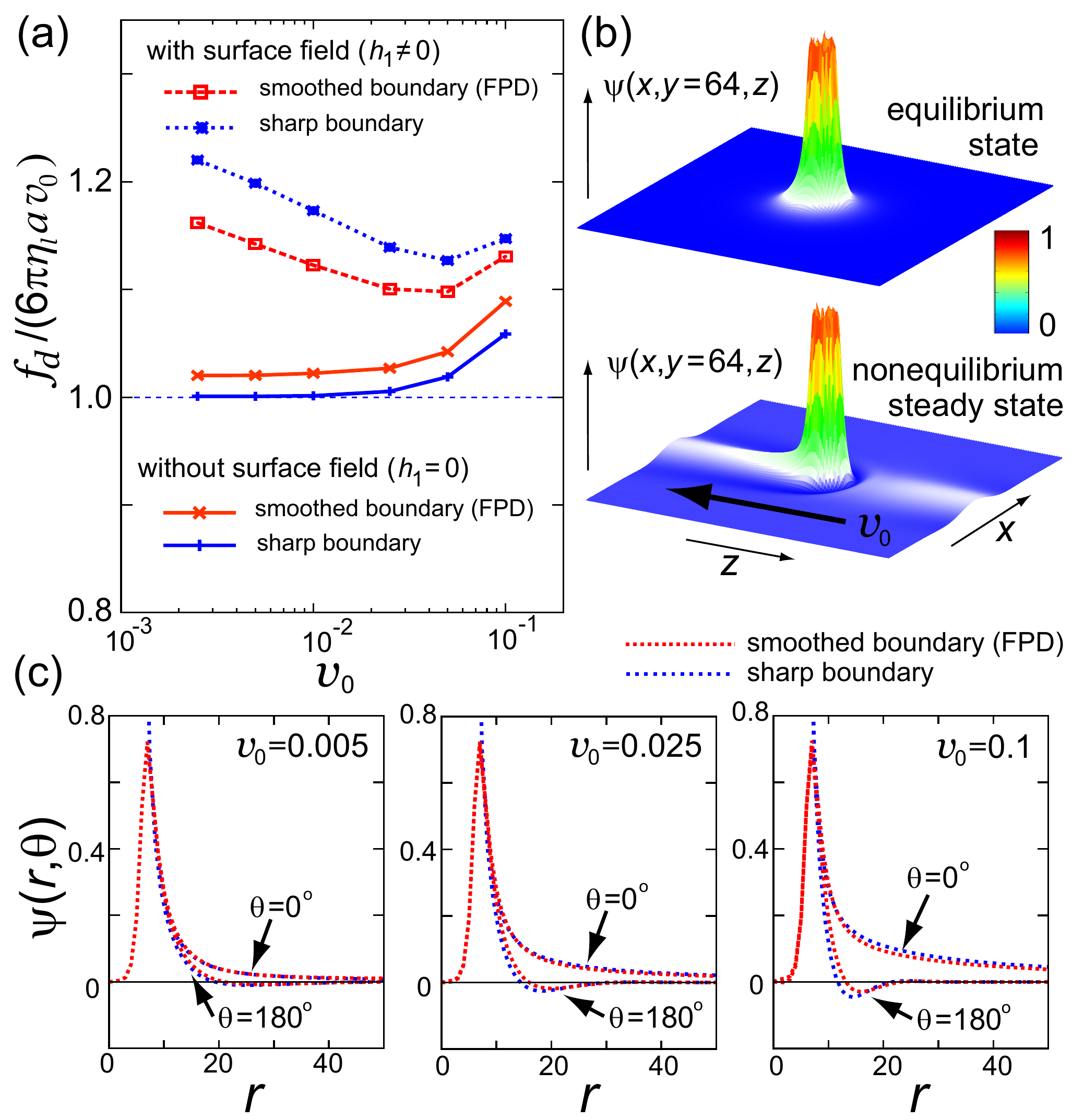}
\end{center}
\caption{(color online) 
(a) The deviation of the drag force from Stokes' law $f_d=6\pi\eta_\ell a v_0$ (dashed horizontal line) as a function of $v_0$ 
for both sharp and smooth boundaries  in the absence (solid curves) or presence (dashed curves) of the surface field. In all cases, ultimately the increase in $v_0$ leads to an increase in the Reynolds number $\mbox{Re}\simeq \rho v_0 a/\eta_\ell$, 
which is concomitant with an increase in viscous dissipation. 
In our units, 
$\mbox{Re} \simeq a v_0$ and deviations from Stokes' drag law are expected to occur for $\mbox{Re} \gtrsim 1$. 
(b) Concentration $\psi$ in the plane $y=64$ generated by a particle with its center fixed at $x=y=z=64$ and with the surrounding fluid either at rest or in motion in negative $z$-direction with far-field velocity $v_0$.
(c) Windward ($\theta=180^\circ$) and leeward ($\theta=0^\circ$) steady-state concentration $\psi$ for various $v_0$ as a function of the 
radial coordinate $r$; $\theta$ is the polar azimuthal angle referred to the direction of $\bm{v_0}$. 
In (a)-(c) $\epsilon=0.025$, $h_1=-4$, $a=6.4$, and $L=1$. }
\label{Fig2}
\end{figure}

\emph{Drag force.---}In order to confront the first non-equilibrium issue, we have calculated the drag force $f_d$ 
(according to Ref.~\cite{furukawa2010}) which has to act on the particle
in order to keep it fixed when exposed to a flow 
with a uniform velocity field ${\bm{v}}_0 = \bm{v}(r\to\infty)$. 
In the sharp boundary case, the stick boundary condition 
${\bm{v}}=0$ is applied at the surface $r=a$ of the particle. 
Figure \ref{Fig2}(a) shows the friction $f_d/(6\pi\eta_\ell a v_0)$ 
for a particle as a function of $v_0$ for both the sharp and the smooth boundary case, with $\epsilon=0.025$
(corresponding, in equilibrium, to $\xi\simeq a$). 
If $h_1 = 0$,  $\psi(\bm{r}) \equiv 0$
due to $T>T_c$ and $\bar \psi=0$. 
Therefore $\psi$ does not affect the velocity field $\bm{v}$, and, as expected,
Stokes' drag law $f_d=6\pi\eta_\ell a v_0$
is reproduced with high accuracy for sufficiently small $v_0$ and sharp boundaries.  
On the other hand, if $h_1\neq  0$ so that $\psi(\bm{r})>0$, $f_d/(6\pi\eta_\ell a v_0)$ increases by
ca.~20\% because there is a drag by the particle and $\psi(\bm{r})$ is  
strongly distorted by the motion (see Fig.~\ref{Fig2}(b)). 
In addition, the friction decreases monotonically upon increasing $v_0$ within the range in which Stokes' law is recovered for $h_1=0$. This trend  is enhanced for larger $\xi$ (not shown), which suggests that the friction is related to the adsorption profile and the concomitant total excess adsorption of the particle.  
As shown in Fig.~\ref{Fig2}(c), $\psi$ around the particle becomes 
increasingly asymmetric with increasing $v_0$, being enhanced leeward and even inverted windward.  
This effect is characterized by the relative importance of advection to diffusion,  
quantified by the P\'eclet number $\textrm{Pe}=\xi v_0/ D_\xi$, 
where $D_\xi$ is the mutual diffusion constant;
within mean-field  theory $D_\xi=L r_0 \epsilon$ in the bulk.
For $\textrm{Pe}\gg 1$ ($v_0\gg 0.004$ in the present analysis),  
the concentration profile around the particle is strongly distorted. 
Reflecting this modification of $\psi$,
the osmotic pressure $\propto \nabla\psi \nabla \psi$ 
acting locally on the particle surface is larger windward    
than leeward, resulting in the total drag force along the $z$-axis. 
The deviation from Stokes' law due to the "dressing" of the particle by $\psi$  
is already a manifestation of non-equilibrium effects.  
We also note that the spatial anisotropy of $\psi$ around the particle implies that 
the CCF between moving particles is also anisotropic. 
The results for the static (Fig.~\ref{Fig1}) and dynamic (Fig.~\ref{Fig2}) behaviors of a single particle also 
demonstrate that the corresponding results with sharp boundaries are approximated well by FPD. 
Therefore it can be reliably employed for studying actual CCFs 
both in equilibrium and non-equilibrium conditions.

\begin{figure}
\begin{center}
\includegraphics[width=0.47\textwidth]{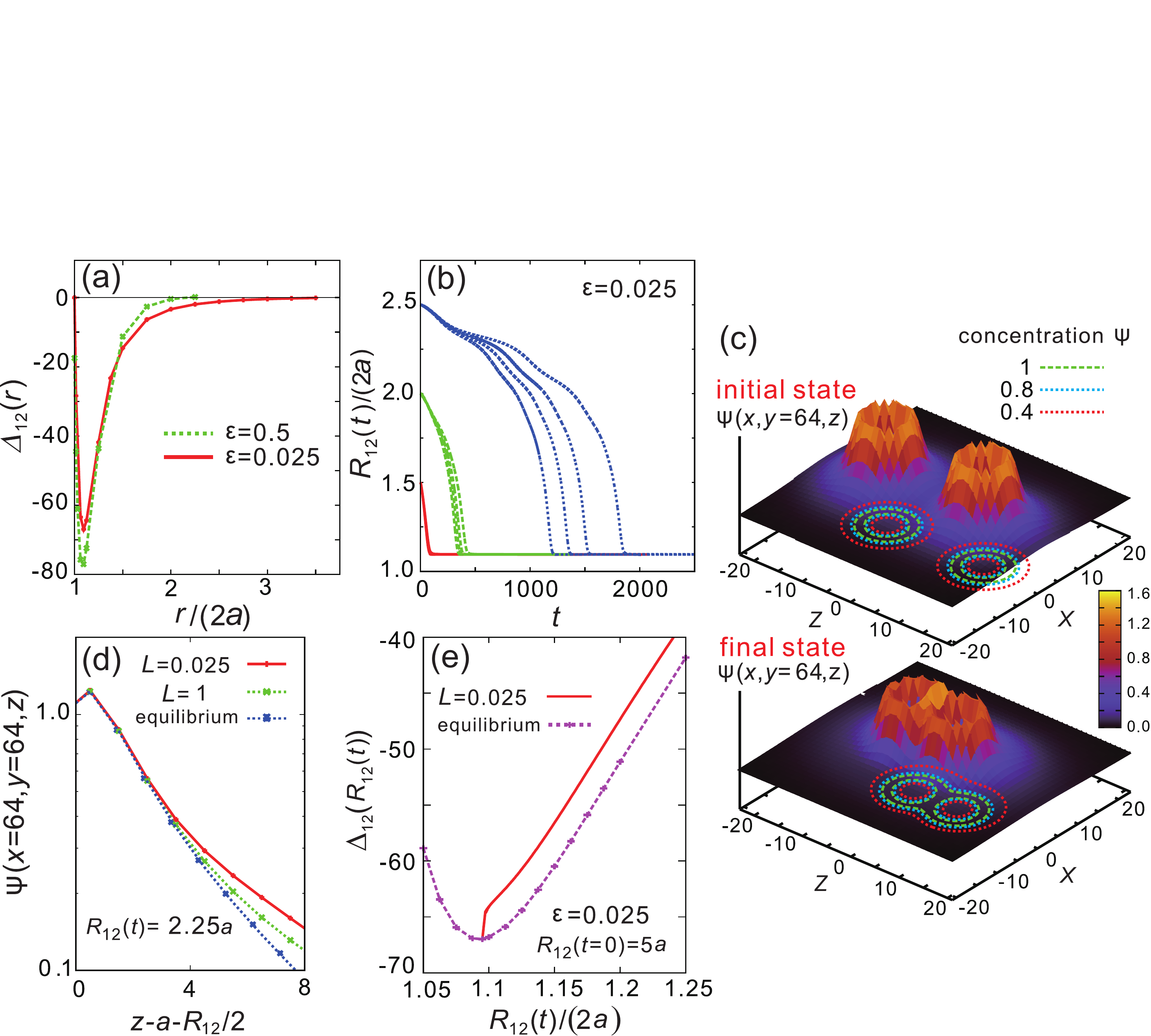}
\end{center}
\caption{(color online)
(a)~Effective interaction energy $\Delta_{\rm 12}(r)$.  
The short-ranged repulsion is due to the direct interaction $u(r)$ and the minimum is located at $r/(2a) \simeq 1.1$.
(b)~Time dependence of the interparticle distance $R_{12}(t)$ 
after releasing the particles at $t=0$ from their 
initial distance $R_{12}(0)$, with an initial equilibrium value of $\psi$.
For $R_{12}(0)=3a$ (red), $4a$ (green), and $5a$ (blue) we consider, from top to bottom, 
$L=0.025$, 0.1, 0.25, and 1. 
(c)~Snapshots of the initial and final concentration profiles and the corresponding contour lines projected onto the $x$-$z$ plane where $\psi=0$ and which contains the centers of the two colloids, located at $z=\pm R_{12}(t)/2$,
for the case $R_{12}(0)=5a$ (uppermost curves in (b)).
The value of  $\psi (x,y=64,z)$ is visualized by the color code. 
(d)~Concentration profiles for the same case as in (c) but at the time when $R_{12}(t)=2.25a$, 
as a function of  $z$ along the axis which joins the centers of the particles. 
(e)~Effective interaction energy (solid line) during the approach of two particles  
as a function of their instantaneous distance, compared with its equilibrium value (dashed line).
In (a)-(e), $h_1=-10$ and $a=4$. 
}
\label{Fig3}
\end{figure}

\emph{Two-particle aggregation.---}With FPD, we now consider two identical colloids 
at center-to-center distance $r$
and we investigate their interaction and the resulting motion.
In order to avoid particle coagulation, we employ $u(r)= (\lambda/r)^{24}E$ with $E/\bar{E}=100$. 
For two reduced temperatures and in equilibrium, Fig.~\ref{Fig3}(a) shows the effective interaction energy $\Delta_{12}(r) = F_{(2)}(r)-2 F_{(1)}$, where $F_{(2)}$ and $F_{(1)}$ are the free energies of a pair and of a single particle, respectively.
Upon approaching $T_c$, one can see the increase in the interaction range, which is set by $\xi$.  
Figure \ref{Fig3}(b) shows the temporal dependence of the interparticle distance $r = R_{12}(t)$
in the process of aggregation during which two particles, originally kept in thermal equilibrium at a distance $r=R_{12}(0)$ and  released at $t=0$, approach each other due to their mutual interaction.
For a fixed value of $R_{12}(0)$, $R_{12}(t)$ strongly depends on the transport coefficient $L$, especially  
if $R_{12}(0)$ is large, indicating that $\psi$ cannot follow the motion of the particle. 
In fact, the smaller the value of $L$, the larger is the corresponding value of $\textrm{Pe}=\xi v_0/(L r_0 \epsilon)$, where $v_0$ is the velocity of approach. 
For two values of $L$, snapshots of $\psi$  
along the direction joining the centers of the particles are shown during the aggregation process and in equilibrium in Fig.~\ref{Fig3}(d).  
As expected, the concentration near the particle surface is almost equilibrated, but far from it,
$\psi$ deviates clearly from the equilibrium profile, 
which becomes more pronounced as $\textrm{Pe}$ increases.

Figure \ref{Fig3}(e) shows the  
evolution of the instantaneous interaction energy %
as function of the interparticle distance during the approach of the two particles
towards the final equilibrium, for which the potential coincides with the equilibrium one (Fig.~\ref{Fig3}(a)). 
During this kinetic process, the interaction strength is always 
weaker than in equilibrium, because in between particles (i.e., in front of a moving particle)
$\psi$ decays on a shorter length scale than its equilibrium counterpart (see Fig.~\ref{Fig2}(b)). 
Therefore the full CCF cannot emerge. 
As expected, this effect is more pronounced for smaller $L$, i.e., larger $\textrm{Pe}$. 
This is a clear signature of non-equilibrium dynamical effects occurring
during the aggregation driven by the attractive CCF, which are enhanced 
as $R_{12}$ decreases because, correspondingly, the driving force and thus the speed of motion
increase. 
For $\textrm{Pe}\gg 1$, the distorted $\psi$ 
around the particles cannot catch up with this motion 
and this retardation leads to strong non-equilibrium effects. This can be clearly seen 
by the very slow reduction of $\Delta_{12}(R_{12}(t))$ even after the two particles have 
reached their final equilibrium positions, corresponding to the almost vertical drop of the solid curve in Fig.~\ref{Fig3}(e). 

\emph{Conclusions.---}The non-equilibrium effects on the CCF investigated here are characterized by the P\'eclet number $\textrm{Pe}=\xi v_0/D_\xi$. For $\textrm{Pe}\gg 1$, 
a large distortion of the concentration field is induced by the particle motion and its extent and anisotropy 
determines the degree of deviation from equilibrium.  
In the presence of thermal fluctuations, 
mode-coupling theory predicts $D_\xi=k_{\rm B}T/(6 \pi \eta_\ell \xi)$ \cite{onuki2002}
where $k_{\rm B}T$ is the thermal energy. The velocity
$v_0$ can be estimated from the steady-state relationship 
$v_0 \simeq {\hat f}/(6 \pi \eta_\ell a)$ where $\hat f$ is the CCF or the drag force acting on the particle. 
Thus $\textrm{Pe} \simeq {\hat f}\xi^2/(a\, k_{\rm B}T)$ and
the situation $\textrm{Pe} \gg 1$ 
($\hat f \gg 30\,$fN, with $\xi \simeq a$ and $\xi \simeq 100\,$nm)   
is within experimental reach \cite{bechinger2008,bechinger2011}. 
Therefore, we expect that significant dynamical effects comparable to those shown in Figs.~\ref{Fig2} and \ref{Fig3} 
should be observable in experiments.  
More detailed analyses including further simulation studies will be presented elsewhere.   
In summary, we have studied non-equilibrium effects on CCFs  
acting on colloids immersed in a near-critical binary liquid mixture.
The effects are intrinsically important and 
sizable near a critical point, and cannot be 
ignored even in the simple case of two particles aggregating under the influence of the CCF themselves.
The non-equilibrium phenomena related to 
CCF are rich and
important for understanding aggregation kinetics of colloidal particles \cite{veen2012},  
the dynamics of a rapidly diffusing colloidal particle, and the interactions between driven colloidal particles  
(e.g., due to diffusiophoresis) near a critical point \cite{buttinoni2012}. 
Similar effects are also expected to occur near an isotropic-nematic 
phase transition of a liquid crystal because of the weak first-order nature of the transition. 
Our FPD method lends itself to addressing these issues and provides an effective tool for investigating novel, 
non-equilibrium features of CCF.

\emph{Acknowledgments.---}This work was partially supported by Grants-in-Aid for Scientific Research (S) and 
Specially Promoted Research from JSPS and by the
Aihara Project, the FIRST program from JSPS, initiated by CSTP. 
H.T.~thanks the Alexander von Humboldt Foundation for kind support. A.G.~was supported by MIUR
within the program ``Incentivazione alla mobilit\`a di studiosi stranieri e italiani residenti all'estero''.

\end{document}